Part 2:

# X-ray reflectometric studies of nanoparticulate hematite films to decouple the rough and smooth behaviors of it and crystallographic and morphological properties concerning fatty acid chain length

Debajeet K. Bora [1, 2, 4*][1], Romy Loehnart [1, 3], Artur Braun[1]

1. Laboratory for High-Performance Ceramics, Empa. Swiss Federal Laboratories of Materials Science and Technology, Uberlandstrasse 129, 8600 Dubendorf, Switzerland
2. Centre for Nano and Material Sciences, JAIN (Deemed - to - be University), Jain Global Campus, Bengaluru, 562112, Karnataka, India
3. Department of SciTec, Ernst- Abbe- Hochschule, 07745 Jena, Germany
4. Laboratory of Inorganic Materials for Sustainable Energy Technologies, University Mohammed VI Polytechnique, Benguerir- 43150, Morocco

**Abstract**

In this study, the use of X-Ray reflectometry technique signifies the types of rough and smooth surfaces of hematite film prepared from different fatty acid derivatives of the iron salt. Followed by this, the film morphology and crystallographic properties concerning different fatty acid chain length have been discussed.

---

[1] Corresponding author: Dr. Debajeet K. Bora, E-Mail: debajeet1@hotmail.com: debajeet.bora@jainuniversity.ac.in

1. **Introduction**

In X-ray reflectometry (XRR) thickness, roughness and density information of single layer and multilayer thin films are determined. The sample is illuminated with a monochromatic X-ray beam at low grazing angles, and the specular x-ray reflectivity is measured. At very low incident angles total reflection of x-rays occurs. For angles bigger than the critical angle of total reflection $\theta_c$ the x-ray beam also penetrates the sample and intensity of reflected x-rays decreases fast with increasing incidence angle. The critical angle depends on the density of the material. For layered samples, the x-rays can be reflected at the inner interfaces and interfere with reflected waves from the surface. Therefore, intensity oscillations, called Kiessig fringes, result in the XRR pattern. The period of fringes gives information about the thickness of layers. The period decreases with higher layer thickness. _Higher surface roughness diminishes specular reflection and leads to a more significant decay of intensity with increasing incidence angle while roughness of interfaces dampens the intensity of Kiessig fringes. XRR is usually used for layer thicknesses from 5Å to 200nm and roughness between 0Å to 20Å [1].

Considering the crystallographic structure of Hematite ($\alpha$-$Fe_2O_3$), it is an iron oxide consisting of hexagonal close-packed oxygen ions and ternary charged iron ions located at two-thirds of the octahedral interstices. It is isostructural with corundum [2] and shows oxygen deficiency [3]. To ensure electroneutrality in $\alpha$-$Fe_2O_3$ sintered at temperatures between 450°C to 800°C oxygen vacancies are compensated by the reduction of $Fe^{3+}$ to $Fe^{2+}$ ions causing n-type semiconducting behavior [3]. Conduction occurs due to hopping of electrons along $Fe^{2+}$-O-$Fe^{3+}$ [4].

A rough surface of the photoelectrode is necessary to achieve good photocurrent values. We have already validated the roughness and thickness of the thin film using profilometric analysis in part 1 of the work. Note that, here a three-layer coated hematite thin film with roughness between 600nm to 800nm and bulk thickness up to 700nm provided photocurrent densities of 0.6mA/$cm^2$. The spin-coated samples were found to be the same in thickness, morphology, crystallographic structure and PEC performance with dip coated ones. The primary motivation of the current investigation is to decipher the surface characteristics of hematite thin film regarding roughness and smoothness and to have a glimpse of crystallographic and morphological properties concerning the change in the chain length of fatty acid or different types of fatty acid.

**2. Materials and methods:** The hematite thin film synthesis and deposition techniques using the fatty acid derivatives of iron salts are already described in part 1 of this work. Here, only the characterization tools and protocols used for studying the reflectometric, crystallographic and morphological properties will be described.

2.1. X-ray reflectometry

For measurement, a Siemens D5000 diffractometer with Cu K$_{\alpha 1}$ radiation was used. To cut off unparallel beam fractions a collimator and LiF-crystal monochromator are installed in front of the detector instead of slits and filters generally used for X-ray diffraction. The sample is fixed to the sample holder by under pressure, and a beam-knife is installed above the sample leaving only a small slot between the sample surface and beam-knife. In this way, the angular divergence of the beam is reduced, and sufficient angular resolution is achieved. Before the measurement, the sample surface has to be aligned precisely to the primary beam direction. By varying the applied voltage and current of the x-ray tube, the intensity of the x-ray beam is adjusted to a maximum count rate of 360000cps in the region of total reflection. The measurement was done in θ/2θ-configuration in a range of $0.07° \leq 2\theta \leq 2°$ with a scan speed of 2s/step and an increment of 0.002°. The intensity of the reflected beam is displayed logarithmical as a function of grazing angle 2θ. The measured curves were normalized to a grazing angle of 0.722° for easier comparison.

2.2. X-Ray Diffraction

The intensity of the diffracted x-rays is detected at an angle 2θ concerning the direction of the incident beam. In the case of constructive interference, the intensity increases significantly. Each family of planes generates an intensity peak at a defined angle θ that is related to the lattice plane distance. From this, the crystal structure of the sample can be identified.

The crystallite size of crystalline material can be calculated using the Scherrer equation. It is based on the broadening of intensity peaks with decreasing crystallite size.

$$d_{XRD} = \frac{K\lambda}{L\cos\theta} \qquad (1)$$

Here K is a shape factor (K=0.89), and L is the value of full width at half maximum of the peak [1].

For measurement, an X'Pert High Score Plus diffractometer by Panalytical with Cu K$_{\alpha 1}$ radiation was used. The sample was fixed and leveled to a sample holder and the intensity measured at an angle 2θ of $5° \leq 2\theta \leq 80°$ with a step size of 0.0167° and scan speed of 0.0857°/s. The measured intensity is displayed as a function of incidence angle 2θ. The phase of hematite and phase purity was verified by comparing the measured XRD pattern with a reference pattern [34]. The crystallite size $d_{XRD}$ was calculated by Scherrer formula using the hematite [104] peak. The exact position of the peak and the value of full width at half maximum were achieved by analysis of the diffractogram using XRD evaluation software X'PERT HighScore Plus.

2.3 Scanning electron microscopy for morphological characterization: For the investigation, a Hitachi S-4800 field emission scanning electron microscope was used. Images based on secondary electrons are taken of the samples using different magnifications from 20000 to 100000 at an acceleration voltage of 10kV to determine the morphology.

3. Results and Discussion

### 3.1. X-ray reflectometric studies of hematite films to decouple the surface characteristics regarding roughness and smoothness

X-ray reflectometry was performed on representative rough one- and two-layer samples of precursor 3 and 4 to further investigate the thickness and roughness of the samples. Here, FTO substrate roughness is too high, which results in the suppression of Kiessig fringes. It can be seen in the XRR pattern of a smooth hematite sample with a thickness of about 60nm and roughness of 25nm (see figure 1a). For comparison in figure 1b, the XRR pattern of a 15nm thick tungsten oxide film with a roughness of 15.1 nm on Titania substrate with the roughness of 1.29nm displays Kiessig fringes. So, conclusions on the hematite layer thickness cannot be made.

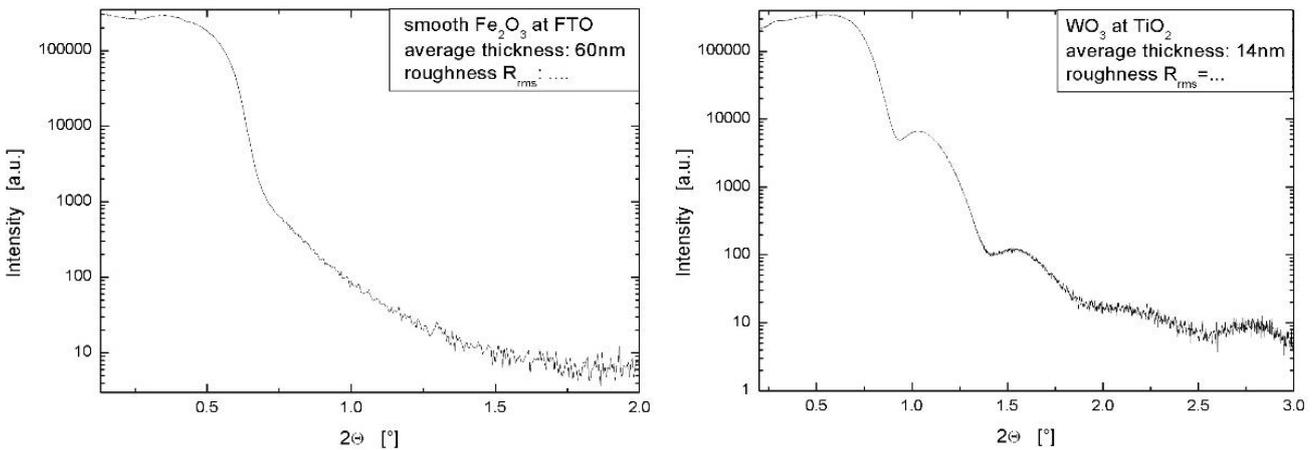

**Figure**1*: a) XRR pattern of a smooth hematite film on FTO substrate, b): XRR pattern of a tungsten oxide film on Titania substrate*

Therefore, the aim is to make a qualitative discussion of the measurements as a higher roughness of the surface increases the decay for angles bigger than the critical angle [1]. Unfortunately, the analysis is found to be sensitive to the adjustment of the slot between the sample and beam knife. Also, different ranges of the roughness of the samples complicate the right alignment. With the bigger size of the slot, the starting intensity of normalized curves is smaller; a shoulder comes up where there is already a decline in intensity for measurements with a lower slot. After this shoulder, the decay starts. However, the slope difference is small for a too big slot and the right size, whereas a too small slot would indicate less roughness because of less decay (see figure 2). For angles *2θ* bigger than 1° there is an intensity difference for the measured samples that also should be due to different adjustments of the slot (see figure3).

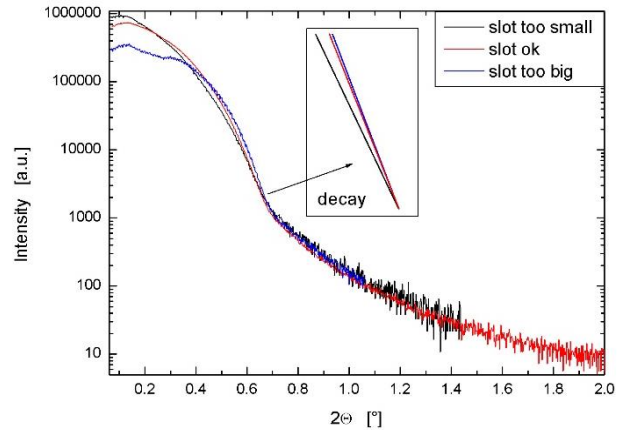

*Figure 2*: *Different XRR decays of one sample due to various changes in the aperture between the sample and wedge*

When comparing the measured curves (see figure 3) the difference in decline could be the difference of roughness of samples, still influenced by the adjustment of the slot. Lines can describe the different decays with the same slope that is taken in the range $0.6°≤2θ≤0.7°$. The decays are qualitatively displayed as a function of the measured value $R_{rms}$ of each sample (see inset figure 3). For each set of samples, a trend of higher decay with bigger $R_{rms}$ can be estimated; it is indicated with a red arrow in the insets. As scattering is very high, it is still to question if the differences in decay represent the roughness of the samples. Slight variations in the adjustment of the slot and differences in alignment of the sample before a measurement could also be the reason for small measurement differences. Furthermore, it is doubtful if the measurement method is sensitive to such high roughnesses as the samples display. Investigation of samples with differently known roughness's on flat substrates should be used to validate the XRR results.

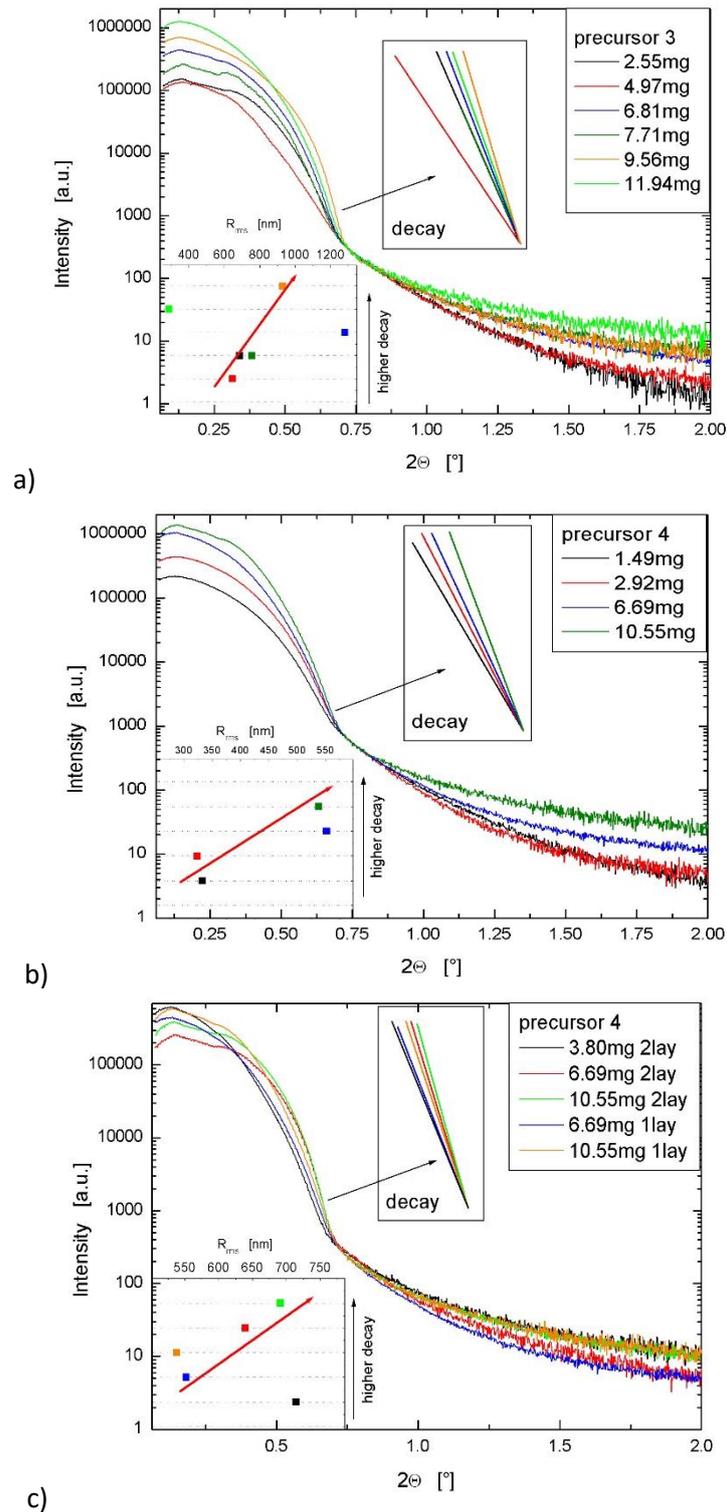

**Figure 3**: *XRR patterns for samples of precursor 3 (a) and one (b) and two (c) layer samples of precursor 4, the red arrows in the insets indicate the increase of roughness of samples with higher decay.*

## 3.2 Crystallographic and morphological properties of nanoparticulate hematite films

*3.2.1. Crystallographic properties of hematite thin films prepared from iron salt oleic acid precursor*

Characteristic rough samples of the weight dependent study and two-layer samples of precursor 3 and precursor four were investigated with x-ray diffraction to determine the crystallographic structure and crystallite size. The XRD patterns also show the tin oxide peaks of the substrate beside the peaks of hematite. No other phase could be identified (see figures 4 and 5). Samples with a higher weight of organic film show slightly increased peak intensity and also for two-layer samples the peak intensity is getting bigger. The crystallite sizes for rough one-layer dip coated samples of precursor three do not differ (see table 1). But for two-layer samples, the size increases slightly. On the contrary one - layer dip coated samples of precursor four show differences in crystallite sizes from 25 nm to 34 nm. By deposition of a second layer, the crystallite size also increases. The photocurrent density of samples of precursor 4 is higher with a bigger crystallite size. Furthermore, a one-layer sample with the same crystallite size of a two-layer sample displays the same current density (see table 2). The increase of crystallite size for two-layer samples can be explained with crystallite growth of the first layer due to additional heat treatment.

*Table 1: Calculated XRD crystallite size of dip coated samples of precursor 3*

| sample weight [mg] | 3.41 | 4.97 | 5.70 | 7.71 | 11.94 | 5.05mg 2 layers |
|---|---|---|---|---|---|---|
| $d_{XRD}$ [nm] | 28.8 | 29.3 | 28.4 | 28.3 | 28.3 | 32.2 |

*Table2: Calculated XRD crystallite size of dip coated samples of precursor 4*

| sample weight [mg] | 1.49 | 2.92 | 3.80 | 6.69 | 10.55 | 6.69 2 layers |
|---|---|---|---|---|---|---|
| $d_{XRD}$ [nm] | 24.7 | 33.7 | 28.6 | 28.8 | 24.8 | 32.4 |
| $J_{ph}$ [$\mu A/cm^2$] | 190 | 334 | 197 | 256 | 187 | 336 |

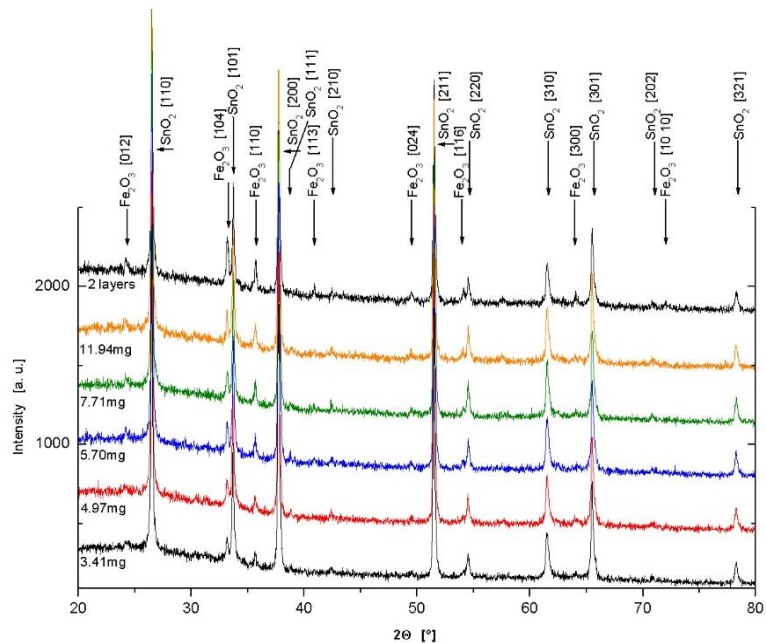

**Figure 4**: *XRD pattern of dip coated samples of precursor 3*

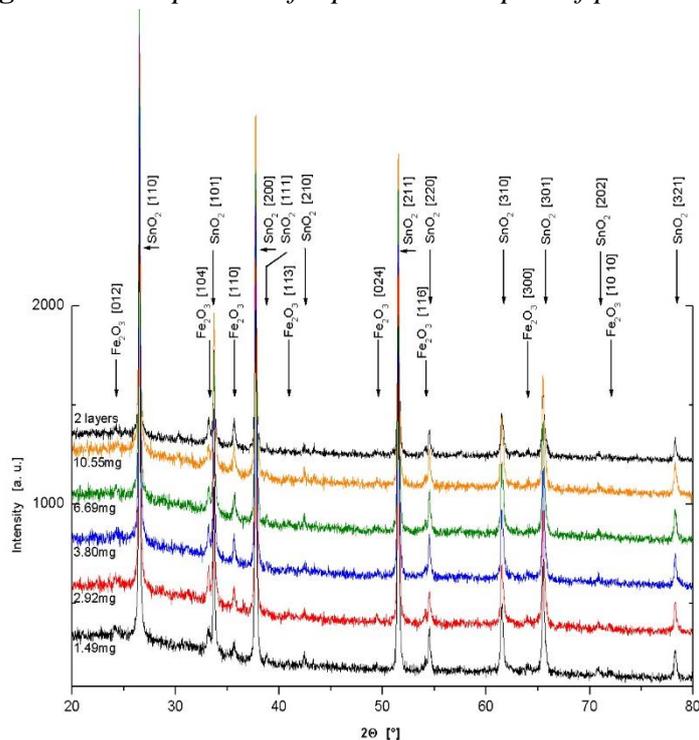

**Figure 1**: *XRD pattern of dip coated samples of precursor 4*

*3.2.2. Morphology of hematite thin films prepared from iron salt oleic acid precursor*

Morphology of dip coated samples was investigated with field emission scanning electron microscopy (FESEM). For rough one-layer samples, the layer thickness is that thin that the topography of substrate (FTO) can mostly be seen (see figure 6). Also, the film is not constant but showing cracks and bare FTO - substrate at some places (see red circle figure 6 a and b). The samples show two different types of morphology. There is a worm-like the structure of particles (orange rectangles figure 6a and 6c) as well as round shaped particles (green circles/ rectangle figure c). They can be either in separate spaces like figure 28 shows or in between the worm-like structure one can find round shaped particles (see figure 6 c). Further investigation of the fraction of single round shaped and single worm like shaped particles could give information about their relevance to the PEC performance. Particle sizes of round shaped particles are from around 70nm of large particles to 20nm for small ones. Samples that show higher photocurrent density also show round shaped particles that display a bigger size, from 90nm up to 250nm (see figure 7).

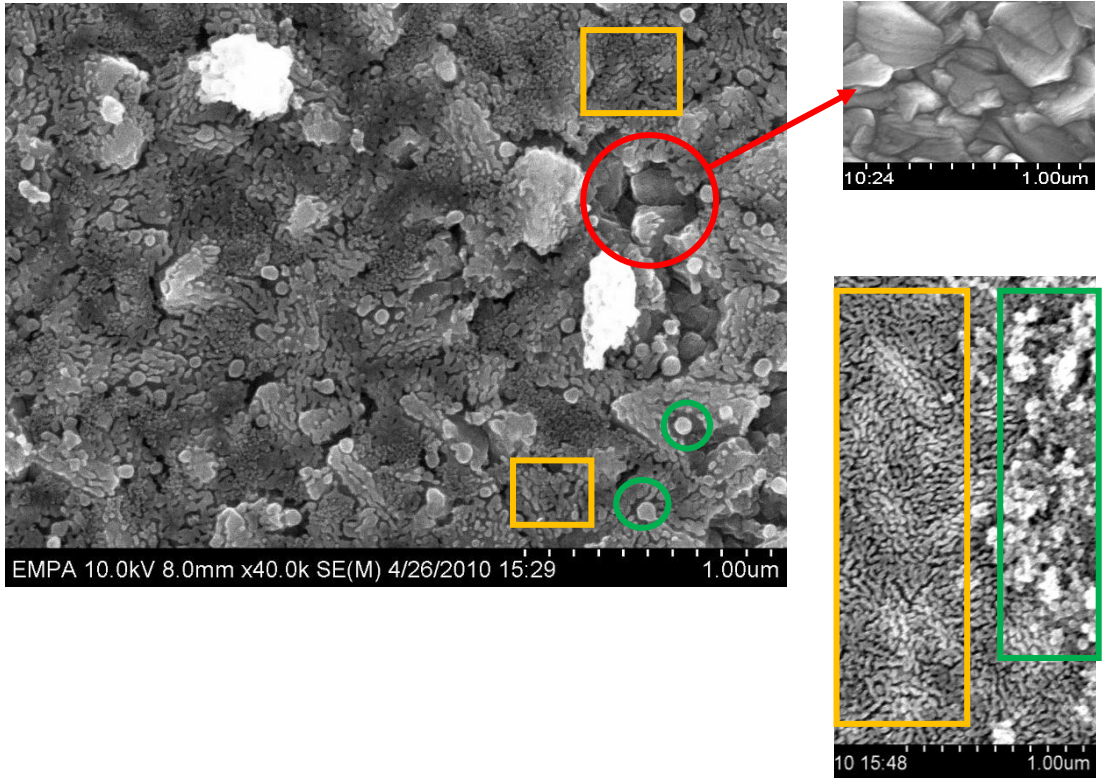

**Figure 6**: *a. SEM Image of hematite film made of iron oleate precursor the green circles and orange rectangles indicate round shaped and worm-like shaped particles, respectively, b. Bare FTO substrate, c. SEM picture is showing worm like shaped (orange rectangle) and round shaped (green box) particles separately.*

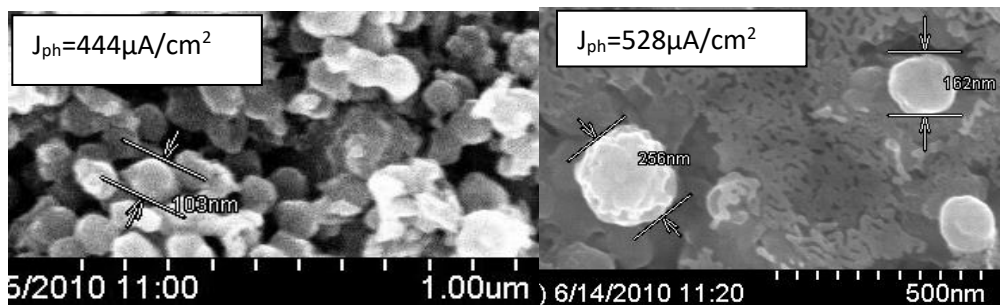

**Figure 7**: *Round shaped particles with a bigger particle size that can be found with samples of higher PEC performance*

A second layer deposited on the rough samples leads to a higher thickness of the film. Therefore, the topography of the substrate cannot be seen anymore. Morphology on the surface does not change, still, there are round shaped and also worm-like particles. For spin-coated samples, the morphology is the same as for dip coated samples, consisting of worm-like and also round formed particles, single and even mixed.

*3.2.3. Crystallographic properties of hematite thin films prepared from iron salt stearic acid precursor*

The XRD - patterns reveal the presence of hematite, no other phase could be identified. Compared to the XRD - trends of samples of iron oleate precursor, the intensity of the hematite phase is higher here. The [110] peak shows higher intensity than the [104] peak, which is the average peak of highest intensity for hematite (see figure 8). Therefore, there should be a preferred orientation of crystals along this crystallographic direction. It has also been described elsewhere for hematite films [5]. The calculated crystallite size for one and two-layer samples is not different, around 24 nm to 26 nm (see table 3).

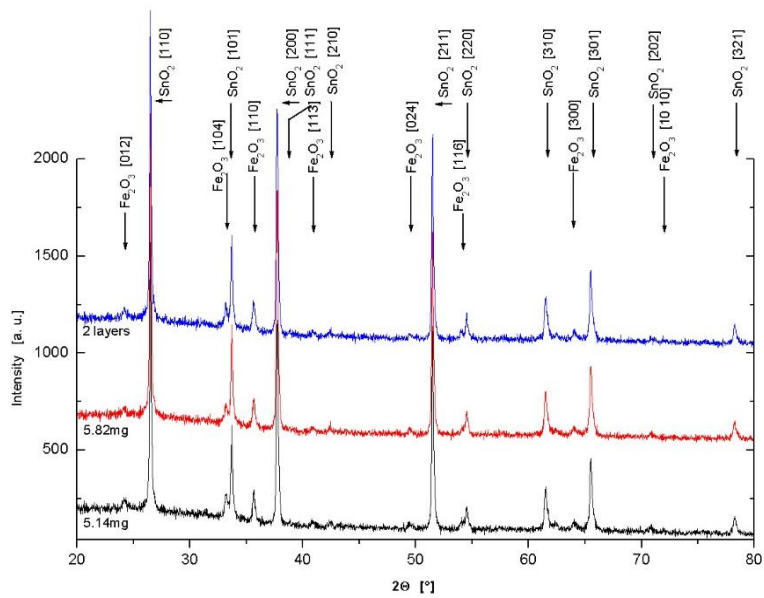

**Figure 8**: *XRD patterns for dip coated samples of stearic acid precursor*

*Table 3: Crystallite size of dip coated samples of stearic acid precursor*

| sample weight [mg] | 5.14 | 5.82 | Two layers |
|---|---|---|---|
| $d_{XRD}$ [nm] | 25.6 | 24.3 | 26.0 |

*3.2.4. Morphology of hematite thin films prepared from iron salt stearic acid precursor*

Morphology of the samples investigated with SEM shows a very thin layer of worm-like shaped particles on FTO substrate for one-layer samples. There are no round shaped particles, the topography of FTO substrate can be seen (see figure 9a). For two-layer samples, the film thickness increases as the substrate topography are not visible anymore (see figure 9b). Still, there is the only worm like shaped particles that seem to emerge from the bulk. They show an aspect ratio between two and three.

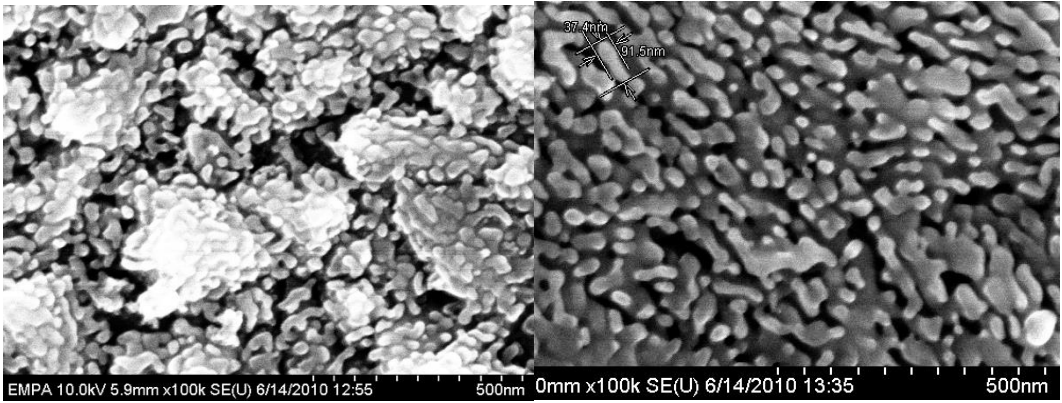

**Figure 2**: SEM picture of one-layer (a) and two layers (b) hematite film made by dip coating of iron stearate precursor

Iron stearate precursor can also be used to prepare photoactive hematite electrodes. Significant differences to hematite films of iron oleate precursor are the preferred orientation of crystals and the absence of round shaped particles. As the reaction of the precursor at preparation was not finished, a more extended heat treatment could affect the PEC performance. Also annealing temperature and dwell time should be optimized to get better information about the PEC performance of hematite samples prepared of this precursor.

*3.2.5. Crystallographic and morphological properties of hematite thin films made from an iron salt lauric acid precursor*

The XRD patterns for one - layer samples reveal the existence of hematite only. Compared to the XRD patterns of samples of oleic acid and stearic acid derivative of the iron precursor, the intensity of hematite peaks is more significant. Also, the intensity of [110] peak found here is higher than the average peak [104] of highest intensity revealing preferred orientation along this crystallographic direction (see figure 10). The crystallite size is in the range of 27 nm to 30 nm (see table 4). Morphology seems to consist of worm-like shaped particles but already in an advanced state of sintering. The structure of a single worm like shaped particles can still be estimated. No round shaped particles were found (see figure 11).

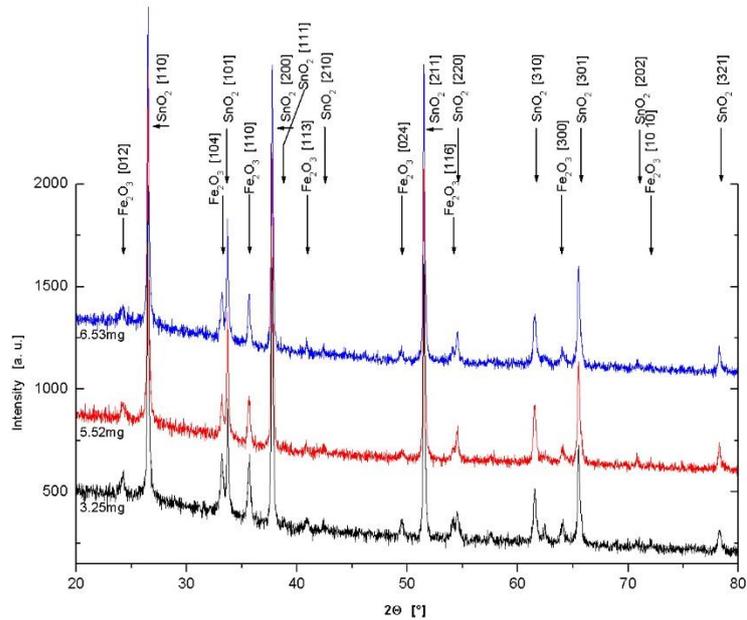

**Figure10**: *XRD patterns for dip coated samples of lauric acid precursor*

*Table 4: Crystallite size for one-layer dip coated samples of lauric acid precursor*

| sample weight [mg] | 3.25 | 5.52 | 6.53 |
|---|---|---|---|
| $d_{XRD}$ [nm] | 26.8 | 29.8 | 28.7 |

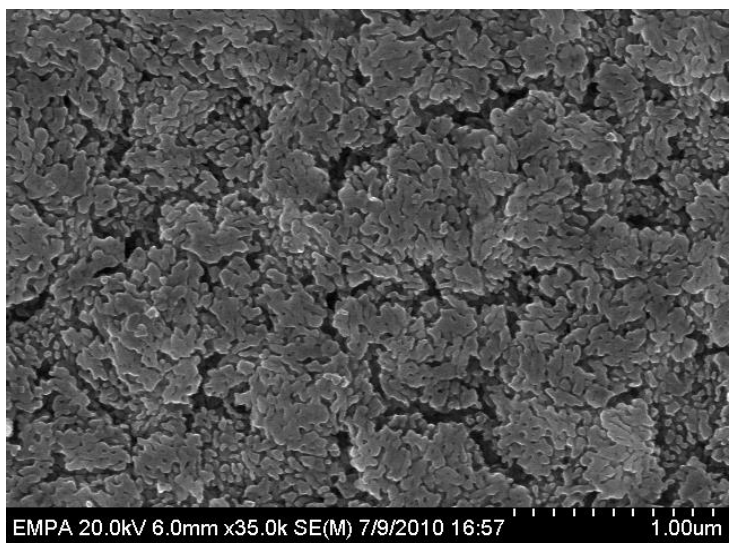

**Figure 11**: *SEM picture of hematite film made by dip coating of iron laurate precursor*

Iron laurate precursor can be used to prepare photoactive hematite electrodes. The samples show a preferred orientation like hematite films prepared of iron stearate precursor do. Morphology also exposes wormlike shaped particles like iron oleate and stearate precursors but more sintered together. Therefore, the investigation concerning dwell time is necessary. Examinations with XRD proved the phase purity of hematite. The crystallite size is nearly the same for the samples, around 30nm. Morphology of samples consists of round shaped and worm-like shaped particles. They can be separated, or the round shaped particles are in between the worm-like shaped particles. Dip and spin coating process show no difference in achieved PEC performance, roughness, thickness or morphology of samples where the only spin coating is more affected regarding less PEC performance by the use of diluted precursor solutions. Films of different precursors prepared with different fatty acids show hematite phase with a preferred orientation along the [110] crystallographic direction. Morphology only reveals a worm-like shaped structure and no round shaped particles. Crystallite size is equal to values of hematite films of iron oleate precursors. Thickness, roughness, and PEC performance are similar to samples of one particular iron oleate precursor.

4. **Conclusion**

Films of different precursors prepared with different fatty acids show hematite phase with a preferred orientation along the [110] crystallographic direction. Morphology only reveals a worm-like shaped structure and no round shaped particles. Crystallite size is equal to values of hematite films of iron oleate precursors. Different precursor compositions had no significant influence on hematite morphology and PEC performance but provided a preferred orientation of hematite [110] crystallographic direction. Reflectometric intensity decline difference signifies that the samples exhibit different roughness's. The roughness of the samples is said to be the main factor for good photocurrent as it allows for a high semiconductor/electrolyte interface,

**Acknowledgment**

Swiss Federal Office of Energy Project No. 100411 and the Swiss National Science Foundation R'Equip No. 206021-121306 to support this research work.